\documentclass[letterpaper, 10 pt, conference]{ieeeconf}  

\IEEEoverridecommandlockouts                              

\usepackage{cite}
\usepackage{amsmath,amssymb,amsfonts}
\usepackage{textcomp}
\usepackage{xcolor}
\usepackage{graphicx}
\usepackage{subfig}
\usepackage{epstopdf}
\usepackage{comment}

\usepackage{amsthm}
\usepackage{enumerate}
\newtheoremstyle{mystl}
  {0}   
  {0}   
  {\normalfont}  
  {\parindent}       
  {\itshape} 
  {:}         
  {5pt plus 1pt minus 1pt} 
  {\thmname{#1} \thmnumber{#2}}          

\theoremstyle{mystl}
\newtheorem{rmk}{Remark}
\newtheorem{lem}{Lemma}
\newtheorem{thme}{Theorem}
\newtheorem{dfn}{Definition}
\newtheorem{prop}{Proposition}
\newtheorem{ass}{Assumption}
\newtheorem{cllry}{Corollary}
\renewcommand{\qed}{\hfill\blacksquare}
\title{\LARGE \bf
Event-triggered Boundary Control of a Class of Reaction-Diffusion PDEs with Time-dependent Reactivity
}
\author{Bhathiya Rathnayake$^{1}$ and Mamadou Diagne$^{2}$
\thanks{$^{1}$B. Rathnayake is with the Department of Electrical and Computer Engineering, University of California San Diego, La Jolla, CA 92093   
        {\tt\small brm222@ucsd.edu}}%
\thanks{$^{2}$M. Diagne is with the Department of Mechanical and Aerospace Engineering, University of California San Diego, La Jolla, CA 92093
        {\tt\small mdiagne@eng.ucsd.edu }}%
}
\begin{document}

\maketitle
\thispagestyle{empty}
\pagestyle{empty}

\begin{abstract}
This paper presents an event-triggered boundary control strategy for a class of reaction-diffusion PDEs with time-varying reactivity under Robin actuation. The control approach consists of a backstepping full-state feedback boundary controller and a dynamic event-triggering condition, which determines the time instants when the control input needs to be updated. It is proved that under the proposed event-triggered boundary control approach, there is a uniform minimal dwell-time between two event times. Furthermore, the well-posedness and the global exponential convergence of the closed-loop system to zero in $L^2$-sense are established. A simulation is conducted to validate the theoretical developments.
\end{abstract}

\section{Introduction}
Event-triggered control is a control technique which updates the control input only at events dictated by a suitable event generator \cite{heemels2012introduction} that utilizes current system states. This is in contrast to conventional sampled-data control \cite{hetel2017recent,karafyllis2017sampled,karafyllis2018sampled} where the control input is updated at certain fixed times periodically or aperiodically irrespective of the system states. Thus, event-triggered control can be viewed as a sampled-data control approach that introduces feedback into communication and control update tasks. In event-triggered control, by exploiting the power of feedback, the control input is updated aperiodically only when needed, which significantly reduces the number of communications and control updates while maintaining a satisfactory closed-loop performance \cite{lemmon2010event}. 

Typically, event-triggered control comprises two main components: a feedback control law that achieves the desired closed-loop performance and an event-triggered mechanism, which determines when to update the control input. A key requirement in event-triggered control is to rule out the so-called \textit{Zeno behavior}, which is the occurrence of infinite number of control updates in a finite time interval. This is typically achieved via careful design of an event triggering mechanism for which there exists a guaranteed lower bound for the time between two successive events, known as the \textit{minimal dwell-time}. 

During the past  years, various interesting results related to event-triggered control have been reported on systems described by both linear and nonlinear ordinary differential equations (ODEs) (see \cite{heemels2012introduction},\cite{liu2015small}). This has kindled a vigorous interest in developing event-triggered control strategies for systems governed by partial differential equations (PDEs) \cite{espitia2016event,espitia2020observer,espitia2019event,katz2020boundary,diagne2021event,rathnayake2021observer,rathnayake2022sampled,wang2021event,wang2022event,rathnayake2022event,rathnayake2022stefan}.   In \cite{espitia2016event}, an output feedback event-triggered boundary controller for 1-D linear hyperbolic systems of conservation laws is proposed using Lyapunov techniques. The authors of \cite{espitia2020observer} design an observer-based event-triggered backstepping boundary controller for a coupled $2\times 2$ hyperbolic system utilizing a dynamic triggering condition. Using ISS properties and small-gain arguments, the authors of \cite{espitia2019event} propose a full-state feedback event-triggered boundary control approach for constant parameter reaction-diffusion PDEs. In \cite{rathnayake2021observer} and \cite{rathnayake2022sampled}, event-triggered boundary control approaches for a class of constant parameter reaction-diffusion PDEs under anti-collocated and collocated boundary sensing and actuation are proposed using dynamic event-triggers. Recently, event-triggered boundary control strategies for the one-phase moving boundary Stefan problem have been proposed under both static \cite{rathnayake2022event} and dynamic \cite{rathnayake2022stefan} triggering conditions.

There are several recent works devoted to the study of boundary control of linear parabolic PDEs with nonconstant parameters \cite{smyshlyaev2005control,meurer2009tracking,kerschbaum2019backstepping,karafyllis2021event}. In \cite{smyshlyaev2005control}, the authors propose backstepping boundary controllers for linear parabolic PDEs with space-dependent diffusivity or time-dependent reactivity. An exponentially stabilizing tracking controller based on backstepping and differential flatness methods is proposed in \cite{meurer2009tracking} for a linear parabolic PDE with spatially and temporally varying coefficients and nonlinear boundary input. The authors of \cite{kerschbaum2019backstepping} develop backstepping boundary controllers for coupled linear parabolic PDEs with space- and time-dependent coefficients. Deviating from \cite{smyshlyaev2005control,meurer2009tracking,kerschbaum2019backstepping}, where one has to solve time varying gain kernel PDEs which capture space- and time-dependent coefficients, the authors of \cite{karafyllis2021event} propose to use a simpler kernel capturing only the spatial variation of the reaction coefficient. This is achieved by sampling the reaction term in time and regarding it constant between two successive sampling times which lead to a time-independent gain kernel PDE. In order to determine these sampling instants, the authors develop static and dynamic event-triggering mechanisms which ensure the global exponential stability of the closed-loop system under continuous-time backstepping boundary control. Despite the reaction term is sampled in time when obtaining the control gain, the control input still has to be continuously computed and applied to the plant. To the best of our knowledge, there are no prior works which address the event-triggered boundary control of infinite-dimensional systems with time-varying parameters.    

In this paper, for the first time, we present an event-triggered boundary control of a class of reaction-diffusion PDEs with time-dependent reactivity under Robin actuation using infinite-dimensional backstepping approach. First we obtain a continuous-time boundary controller which we apply in a zero-order hold fashion between two sampling instants to obtain the event-triggered control input. We develop a dynamic event-triggering mechanism to determine the sampling instants when the control input needs to be computed and updated. Under the proposed event-triggering mechanism, we prove that the closed-loop system is Zeno-free. Furthermore, we establish the closed-loop system's well-posedness and the global exponential convergence to zero in $L^2$-sense. 

This paper is organized as follows. Section II introduces the class of linear reaction-diffusion system and the continuous-time event-triggered boundary control. Section III presents the event-triggered boundary control. A numerical exampled is provided in Section IV to illustrate the results, and the conclusion is provided in Section V.

\textit{Notation:} $\mathbb{R}_{+}$ is the nonnegative real line whereas $\mathbb{N}$ is the set of natural numbers including zero.  By $C^{0}(A;\Omega)$, the class of continuous functions on $A\subseteq\mathbb{R}^{n}$ is denoted, which takes values in $\Omega\subseteq\mathbb{R}$. By $C^{k}(A;\Omega)$, where $k\geq 1$, the class of continuous functions on $A$, which takes values in $\Omega$ and has continuous derivatives of order $k$, is denoted.  $L^{2}(0,1)$ stands for the equivalence class of Lebesgue measurable functions $f:[0,1]\rightarrow\mathbb{R}$ such that $\Vert f\Vert=\big(\int_{0}^{1}\vert f(x)\vert^{2}\big)^{1/2}<\infty$. Let $u:[0,1]\times\mathbb{R}_{+}\rightarrow\mathbb{R}$ be given. $u[t]$ represents the profile of $u$ at certain $t\geq 0$, \textit{i.e.,} $\big(u[t]\big)(x)=u(x,t),$ for all $x\in [0,1]$. For an interval $J\subseteq\mathbb{R}_{+},$ the space $C^{0}\big(J;L^{2}(0,1)\big)$ is the space of continuous mappings $J\ni t\rightarrow u[t]\in L^{2}(0,1)$.  

\section{Continuous-time Backstepping Boundary Control}
Let us consider the following scalar reaction-diffusion system:
\begin{subequations}\label{ctp}
\begin{align}\label{ctpe1}
u_{t}(x,t)&=\varepsilon u_{xx}(x,t)+\lambda(t) u(x,t),\\
\label{ctpe2}
u_{x}(0,t)&=0,\\\label{ctpe3}
u_{x}(1,t)+qu(1,t)&=U(t),
\end{align}
\end{subequations}
and the initial condition $u[0]\in L^{2}(0,1),$ where $\varepsilon>0$ is a constant, $u: [0,1]\times [0,\infty)\rightarrow\mathbb{R}$ is the system state and $U(t)$ is the control input. The time-dependent reaction coefficient $\lambda(t)$ satisfies the following assumption:

\medskip

\begin{ass}\label{ass1}
The reaction coefficient $\lambda(t)$ satisfies
\begin{equation} \label{xx1}
    \sup_{t\geq 0}\vert \lambda^{(n)}(t)\vert\leq D^{n+1}n!,
\end{equation}
for all $t\geq 0,$ where $D>0$ is a constant. 
\end{ass}

\medskip

\begin{ass}\label{ass2}
 The plant’s parameters $q$ and $\varepsilon$ satisfy the following inequality
 \begin{equation}
    q > \frac{D +\varepsilon}{2\varepsilon}.
\end{equation}
\end{ass}

\medskip

\begin{rmk}\label{rem1} Assumption 1 assumes  $\lambda(t)$ belongs to Gevrey class $G_{M,R,\alpha}(\mathbb{R}_+)$ of order $\alpha= 1$ in $\mathbb{R}_+$ with $M=D$ and $R=1/D$ \cite{meurer2009tracking}. This makes $\lambda(t)$ an analytic function of $t$. Further, Assumption \ref{ass1} is required to ensure the boundedness of the solution of the time-varying gain kernel and its certain spatial and time derivatives, which emanate from the backsteping (inverse) transform.  Assumption \ref{ass2} is required to avoid a trace term for which it is impossible to obtain a useful bound on its rate of change. In order to overcome this, it is required that $q-D/2\varepsilon$ is sufficiently large. 
\end{rmk}

\medskip 

\begin{prop}\label{prop2} The invertible backstepping transformation
\begin{equation}\label{ctbt}
w(x,t)=u(x,t)-\int_{0}^{x}K(x,y,t)u(y,t)dy,
\end{equation}
where $K(x,y,t)$ satisfies
\begin{subequations}\label{ctck}
\begin{align}\label{ctcke1}
\begin{split}
K_t(x,y,t)&=\varepsilon K_{xx}(x,y,t)-\varepsilon K_{yy}(x,y,t)\\&\quad-\lambda(t)K(x,y,t),
\end{split}
\\
K_y(x,0,t)&=0,
\\\label{eew}
K(x,x,t)&=-\frac{\lambda(t)}{2\varepsilon}x,
\end{align}
\end{subequations}
for $0\leq y\leq x\leq 1,\text{ and } t\geq 0,$  and a control law $U(t)$ chosen as\begin{equation}\label{ctcl}
U(t)=\int_{0}^{1}\Big(r(t)K(1,y,t)+K_{x}(1,y,t)\Big)u(y,t)dy,
\end{equation}map the system \eqref{ctp} into the following target system:
\begin{subequations}\label{etots}\begin{align}\label{etotse1}
w_{t}(x,t)&=\varepsilon w_{xx}(x,t),
\\\label{etotse2}
w_x(0,t)&=0,
\\\label{etotse3}
w_{x}(1,t)&=-r(t)w(1,t),
\end{align}\end{subequations}
where
\begin{equation}\label{RR}
    r(t)=q-\frac{\lambda(t)}{2\varepsilon}. 
\end{equation}
\end{prop}

\medskip

The inverse transformation of \eqref{ctbt} is given by \begin{equation}\label{ink}
u(x,t)=w(x,t)+\int_{0}^{x}L(x,y,t)w(y,t)dy,
\end{equation}
where $L(x,y,t)$ satisfies 
\begin{subequations}\label{ctck2}
\begin{align}\label{ctcke11}
\begin{split}
L_t(x,y,t)&=\varepsilon L_{xx}(x,y,t)-\varepsilon L_{yy}(x,y,t)\\&\quad+\lambda(t)L(x,y,t),
\end{split}
\\
L_y(x,0,t)&=0,
\\\label{eew11}
L(x,x,t)&=-\frac{\lambda(t)}{2\varepsilon}x,
\end{align}
\end{subequations}
for $0\leq y\leq x\leq 1,\text{ and } t\geq 0$. 

\medskip

\begin{rmk} The explicit solution to the gain kernels \eqref{ctck} and \eqref{ctck2} are respectively given by \begin{equation}\label{exK}
    \begin{split}
    K(x,y,t)=&-\frac{x}{2}e^{-\int_{0}^t\lambda(\xi)d\xi}\\&\times\sum_{n=0}^{\infty}\frac{1}{n!(n+1)!}\Big(\frac{x^2-y^2}{4\varepsilon}\Big)^n F^{(n)}(t),
    \end{split}
\end{equation} where
\begin{equation}\label{ggh}
  F(t)=\frac{\lambda(t)}{\varepsilon}e^{\int_{0}^t\lambda(\xi)d\xi},
\end{equation}
and 
\begin{equation}\label{exK1}
    \begin{split}
    L(x,y,t)=&\frac{x}{2}e^{\int_{0}^t\lambda(\xi)d\xi}\\&\times\sum_{n=0}^{\infty}\frac{1}{n!(n+1)!}\Big(\frac{x^2-y^2}{4\varepsilon}\Big)^n G^{(n)}(t),
    \end{split}
\end{equation}
where
\begin{equation}\label{gghb}
  G(t)=-\frac{\lambda(t)}{\varepsilon}e^{-\int_{0}^t\lambda(\xi)d\xi}.
\end{equation}
In Appendix A, we provide their derivations. The solutions \eqref{exK} and \eqref{exK1} are rather explicit. Since $x^2-y^2\leq 1$, one can obtain very accurate
approximations to $K(x,y,t)$ and $L(x,y,t)$ by using just several terms of the
sums in \eqref{exK} and \eqref{exK1}, respectively.
\end{rmk}

\subsection{Emulation of the backstepping boundary control law}
We aspire to stabilize the system \eqref{ctp} while sampling the continuous-time controller $U(t)$ given by \eqref{ctcl} at a certain sequence of time instants $(t_{j})_{j\in\mathbb{N}}$. These time instants will be precisely characterized later on based on a dynamic event trigger. The control input is held constant between two successive time instants and is updated when a certain condition is met. Therefore, we define the control input for $t\in[t_{j},t_{j+1}),j\in\mathbb{N}$ as
\begin{equation}\label{etcl}
U_{j}:=U(t_{j})=\int_{0}^{1} k(y,t_j) u(y,t_{j})dy,
\end{equation}
where 
\begin{equation}\label{kyt}
    k(y,t)=r(t)K(1,y,t)+K_{x}(1,y,t).
\end{equation}
Accordingly, the boundary condition \eqref{ctpe3} is modified as follows:\begin{equation}\label{mctpe3}
u_{x}(1,t)+qu(1,t)=U_{j}.
\end{equation}
The deviation between the continuous-time control law and its sampled counterpart, referred to as the input holding error, is defined as follows:
\begin{equation}\label{ffty}
    d(t):= U_j-U(t)=\int_{0}^1\big(k(y,t_j) u(y,t_j)-k(y,t) u(y,t)\big)dy,
\end{equation}
for $t\in[t_j,t_{j+1})$. It can be shown that the backstepping transformation \eqref{ctbt},\eqref{ctck} and the sampled-data control input \eqref{etcl} applied on the system \eqref{ctpe1},\eqref{ctpe2},\eqref{mctpe3} between $t_{j}$ and $t_{j+1}$, yield the following target system, valid for $t\in[t_{j},t_{j+1}),j\in\mathbb{N}$
\begin{subequations}\label{ettsm}\begin{align}\label{ettsm1}
w_{t}(x,t)&=\varepsilon w_{xx}(x,t),
\\\label{ettsm2}
w_x(0,t)&=0,
\\\label{ettsm3}
w_{x}(1,t)&=-r(t)w(1,t)+d(t),
\end{align}\end{subequations}
where $r(t)$ is given by \eqref{RR}. 
\subsection{Well-posedness issues}

\medskip

\begin{thme}\label{thm11}
Under Assumption \ref{ass1}, for every $u[t_j]\in L^2(0,1)$ and constant $U_j\in\mathbb{R}$, there exists a unique solution $u:[t_j,t_{j+1}]\times[0,1]\rightarrow\mathbb{R}$ between two time instants $t_{j}$ and $t_{j+1}$ such that $u\in C^0\big([t_j,t_{j+1}];L^2(0,1)\big)\cap C^1\big((t_j,t_{j+1})\times [0,1]\big)$ with $u[t]\in C^2([0,1])$  which satisfy \eqref{ctpe2},\eqref{mctpe3} for $t\in(t_j,t_{j+1}]$ and \eqref{ctpe1} for $t\in(t_j,t_{j+1}],x\in(0,1)$.   
\end{thme}

\textit{Proof}: Due to Assumption \ref{ass1}, we have that 
\begin{equation}
    \vert\lambda(t)\vert\leq D,
    \end{equation}
and
\begin{equation}
    \vert \dot{\lambda}(t)\vert\leq D^2,
\end{equation}
for all $t\geq 0$. Since the first derivative is bounded, the following Lipschitz condition holds
    \begin{equation}
        \vert \lambda(t)-\lambda(s)\vert \leq D^2\vert t-s\vert,
    \end{equation}
for all $t\geq 0$. Further, let 
\begin{equation}\label{yyt}
    v(x,t):=u(x,t)-\frac{U_j}{q}.
\end{equation}
Then, we have that
\begin{align}\label{v1}
    v_t(x,t)&=\varepsilon v_{xx}(x,t)+\lambda(t)\Big(v(x,t)+\frac{U_j}{q}\Big),\\\label{v2}
    v_x(0,t)&=0,\\\label{v3}
    v_x(1,t)+qv(1,t)&=0.
\end{align}
As $u[t_j]\in L^2(0,1)$, we can verify from \eqref{yyt} that $v[t_j]\in L^2(0,1)$. Thus, from straightforward application of Theorem 2.2 in \cite{karafyllis2021event}, we can obtain that there exists a unique solution
$v:[t_j,t_{j+1}]\times[0,1]\rightarrow\mathbb{R}$ between two time instants $t_{j}$ and $t_{j+1}$ such that $v\in C^0\big([t_j,t_{j+1}];L^2(0,1)\big)\cap C^1\big((t_j,t_{j+1})\times [0,1]\big)$ with $v[t]\in C^2([0,1])$  which satisfy \eqref{v2},\eqref{v3} for $t\in(t_j,t_{j+1}]$ and \eqref{v1} for $t\in(t_j,t_{j+1}],x\in(0,1)$. Therefore, considering \eqref{yyt}, we can conclude that the system \eqref{ctpe1},\eqref{ctpe2},\eqref{mctpe3} has a unique solution between two time instants $t_j$ and $t_{j+1}$ in the sense of Theorem \ref{thm11}.

\section{Event-triggered Boundary Control}

\begin{figure}
\centering
\includegraphics[scale=0.525]{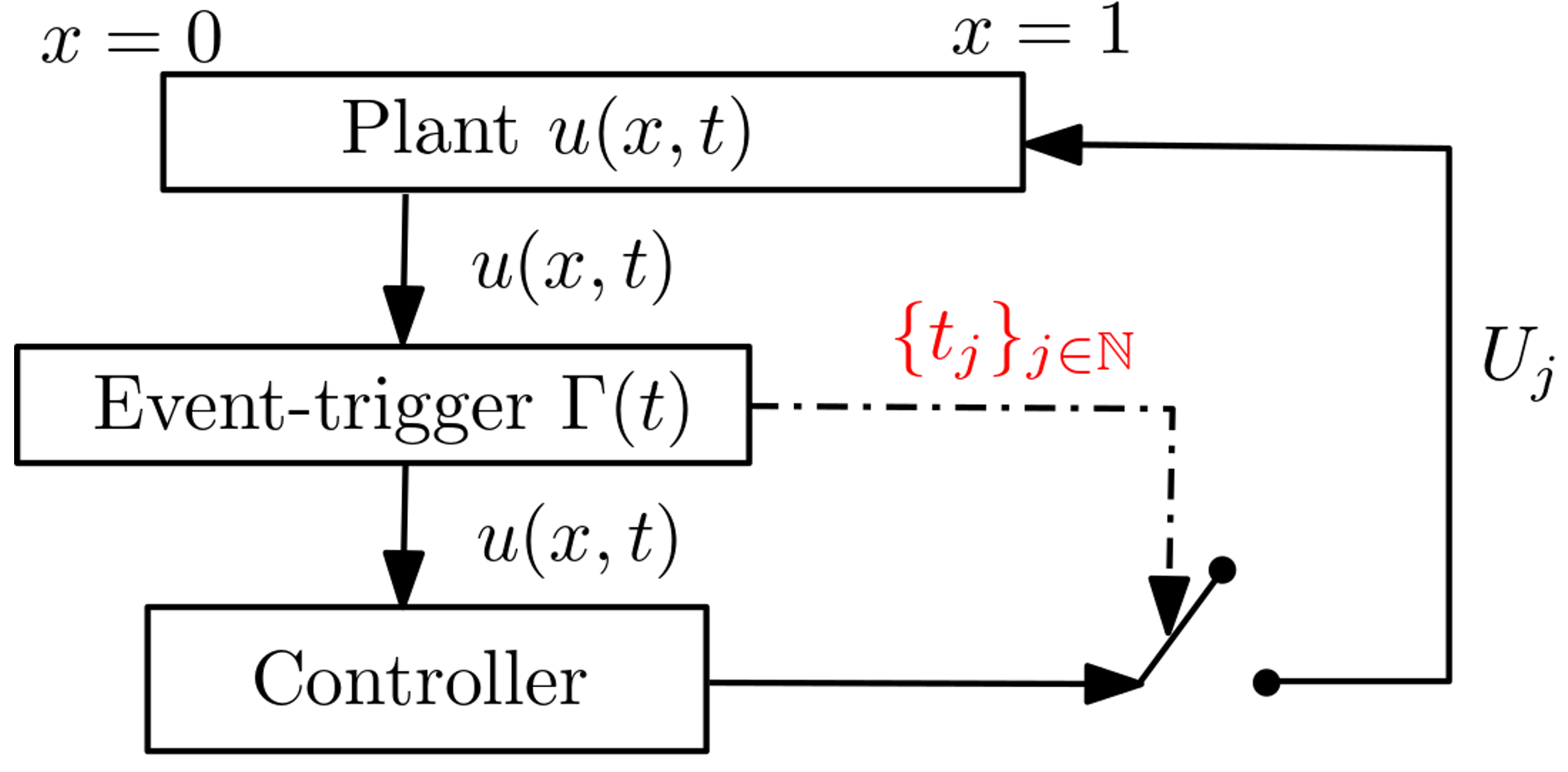}
\caption{Event-triggered closed-loop system.}
\end{figure}
Let us now present the event-triggered boundary control approach considered in this work. It consists of two components: 1) a dynamic event-triggering mechanism which decides the time instants at which the control value needs to be sampled/updated and 2) the infinite-dimensional backstepping boundary control applied in  a Zero-order hold fashion. The structure of the closed-loop system consisting of the plant, controller, and the event trigger is illustrated in Fig. 1.

\begin{dfn}\label{def1}
Let $\eta,\gamma,\rho,\beta_{1},\beta_{2}>0$. The event-triggered boundary control strategy consists of two components.
\begin{enumerate}
 \item (The control action) The event-triggered boundary control input\begin{equation}\label{obetbc4}
U_{j}=\int_{0}^{1}k(y,t_j)u(y,t_{j})dy,
\end{equation}for all $t\in[t_{j},t_{j+1}),j\in\mathbb{N}$ where $k(y,t)$ is given by \eqref{kyt}.

\item (The event-trigger) The set of event times $I=\{t_0,t_1,t_2,\ldots\}$ with $t_0=0$ forms an increasing sequence via the following rule: 
 \begin{equation}\label{obetbc1}
t_{j+1}=\inf\{t\in\mathbb{R}_{+}|t>t_{j}, d^{2}(t)>\gamma m(t)\}
\end{equation}  and $m(t)$ satisfies the ODE 
\begin{equation}\label{obetbc3}\begin{split}
\dot{m}(t)=&-\eta m(t)-\rho d^{2}(t)+\beta_{1}\Vert u[t]\Vert^{2}+\beta_{2}\vert u(1,t)\vert^{2},\end{split}
\end{equation}  for all $t\in(t_{j},t_{j+1})$ with $m(t_{0})=m(0)>0$ and $m(t_{j}^{-})=m(t_{j})=m(t_{j}^{+})$.  
\end{enumerate}
\end{dfn}

\medskip

\begin{lem}\label{pom}Under the definition of the event-trigger \eqref{obetbc1}-\eqref{obetbc3}, it holds that $d^2(t)\leq\gamma m(t)$ and $m(t)>0$, for all $t\in[0,\sup(I))$.
\end{lem}
\textit{Proof:} The proof is very similar to that of Lemma 1 in \cite{rathnayake2021observer}, and hence omitted. \hfill$\qed$

\medskip

\begin{lem}\label{ddf1} For $F(t)$ given by \eqref{ggh}, it holds that 
\begin{equation}\label{bbn}
\begin{split}
    F^{(n)}(t)=&\frac{\lambda^{(n)}(t)}{\varepsilon}e^{\int_{0}^{t}\lambda(\xi)d\xi}\\&+\sum_{m=1}^{n}\frac{n!}{(n-m)!m!}\lambda^{(n-m)}(t)F^{(m-1)}(t),
\end{split}
\end{equation}
for any $n\in\mathbb{N}\text{\textbackslash}\{0\}$. Further, let $\lambda(t)$ satisfy Assumption \ref{ass1}. Then, it holds that 
    \begin{equation}\label{vvm}
    \vert F^{(n)}(t)\vert\leq \frac{(n+1)! D^{n+1}}{\varepsilon}e^{\int_{0}^t\lambda(\xi)d\xi},
\end{equation}
for any $n\in\mathbb{N}$. 
\end{lem}

\textit{Proof:} From mathematical induction, we can easily show that \eqref{bbn} holds for all $n=1,2,\ldots$. For $n=0$, due to Assumption \ref{ass1}, we have that 
\begin{equation}
    \vert F^{(0)}(t)\vert=\vert F(t)\vert\leq \frac{D}{\varepsilon} e^{\int_{0}^{t}\lambda(\xi)d\xi},
\end{equation}
and thus, relation \eqref{vvm} holds for $n=0$. For any $n\in\mathbb{N}\text{\textbackslash}\{0\}$, we can write from \eqref{bbn} that 
\begin{equation}\label{hhjm}
\begin{split}
    \vert F^{(n)}(t)\vert\leq &\frac{\vert\lambda^{(n)}(t)\vert}{\varepsilon}e^{\int_{0}^{t}\lambda(\xi)d\xi} \\&+\sum_{m=1}^{n}\frac{n!}{(n-m)!m!}\vert\lambda^{(n-m)}(t)\vert\cdot\vert F^{(m-1)}(t)\vert.
\end{split}
\end{equation}
Let us assume that the relation \eqref{vvm} holds for all $n=1,\ldots,n-1$. Then, using Assumption \ref{ass1}, we can obtain from \eqref{hhjm} that 
\begin{equation}
\begin{split}
    \vert F^{n}(t)\vert \leq &\frac{D^{n+1}n!}{\varepsilon}e^{\int_{0}^{t}\lambda(\xi)d\xi}+\frac{D^{n+1}}{\varepsilon}e^{\int_{0}^{t}\lambda(\xi)d\xi}\sum_{m=1}^{n}n!\\\leq & \frac{D^{n+1}n!}{\varepsilon}e^{\int_{0}^{t}\lambda(\xi)d\xi}+n\frac{D^{n+1}n!}{\varepsilon}e^{\int_{0}^{t}\lambda(\xi)d\xi}
    \\\leq &\frac{(n+1)!D^{n+1}}{\varepsilon}e^{\int_{0}^{t}\lambda(\xi)d\xi},
\end{split}
\end{equation}
which completes the proof. \hfill$\qed$
\medskip

\begin{lem}\label{gggk} Let the solution to the  gain kernel equations \eqref{ctcke1}-\eqref{eew} be given by \eqref{exK}. Then, subject to Assumption \ref{ass1}, the following bounds can be obtained 
 for all $0\leq y\leq x\leq 1$ and $t\geq 0$.:
    \begin{align}\label{ww1}
        \vert K(x,y,t)\vert&\leq \frac{D}{2\varepsilon} e^{\frac{D}{4\varepsilon}},\\\label{ww2}
        \vert K_t(x,y,t)\vert &\leq \frac{D^2}{2\varepsilon}\Big(3+\frac{D}{4\varepsilon}\Big)e^{\frac{D}{4\varepsilon}},\\\label{ww3}
        \vert K_x(x,y,t)\vert&\leq \frac{D}{2\varepsilon}\Big(1+\frac{D}{2\varepsilon}\Big)e^{\frac{D}{4\varepsilon}},\\
        \vert K_{xt}(x,y,t)\vert&\leq \frac{D^2}{\varepsilon}\Big(\frac{3}{2}+\frac{9D}{8\varepsilon}+\frac{D^2}{16\varepsilon^2}\Big)e^{\frac{D}{4\varepsilon}},\\
        \vert K_{xy}(x,y,t)\vert&\leq \frac{D^2}{4\varepsilon^2}\Big(1+\frac{D}{2\varepsilon}\Big)e^{\frac{D}{4\varepsilon}},\\
        \vert K_y(x,y,t)\vert&\leq \frac{D^2}{4\varepsilon^2} e^{\frac{D}{4\varepsilon}},\\
        \vert K_{yy}(x,y,t)\vert&\leq\frac{D^2}{4\varepsilon^2}\Big(1+\frac{D}{2\varepsilon}\Big)e^{\frac{D}{4\varepsilon}},\\\label{ww8}
        \vert K_{xyy}(x,y,t)\vert&\leq \frac{D^2}{4\varepsilon^2}\Big(1+\frac{D}{2\varepsilon}+\frac{D^2}{4\varepsilon^2}\Big)e^{\frac{D}{4\varepsilon}},
    \end{align}
\end{lem}

\textit{Proof:} 
Due to page limitations, we will only provide the derivations of the relations \eqref{ww1}-\eqref{ww3}. The rest of the results can be obtained by following the same arguments used to derive \eqref{ww1}-\eqref{ww3}. 

Using \eqref{exK}, relation \eqref{vvm} in Lemma \ref{ddf1}, and the fact that $0\leq y\leq x\leq 1$, we can obtain that 
\begin{equation}
\begin{split}
    \vert K(x,y,t)\vert &\leq\frac{D}{2\varepsilon}\sum_{n=0}^{\infty}\frac{1}{n!}\Big(\frac{(x^2-y^2)D}{4\varepsilon}\Big)^n\\&\leq 
    \frac{D}{2\varepsilon} e^{\frac{(x^2-y^2)D}{4\varepsilon}}\leq \frac{D}{2\varepsilon} e^{\frac{D}{4\varepsilon}}.
\end{split}
\end{equation}
\medskip
Differentiating \eqref{exK} with respect to $t$, we can obtain that
\begin{equation}
\begin{split}
    &K_t(x,y,t)\\&=\frac{x}{2}\lambda(t)e^{-\int_{0}^t\lambda(\xi)d\xi}\sum_{n=0}^{\infty}\frac{1}{n!(n+1)!}\Big(\frac{x^2-y^2}{4\varepsilon}\Big)^n F^{(n)}(t)\\&-\frac{x}{2}e^{-\int_{0}^t\lambda(\xi)d\xi}\sum_{n=0}^{\infty}\frac{1}{n!(n+1)!}\Big(\frac{x^2-y^2}{4\varepsilon}\Big)^{n} F^{(n+1)}(t).
\end{split}
\end{equation}
Using Assumption \ref{ass1}, relation \eqref{vvm} in Lemma \ref{ddf1}, and the fact that $0\leq y\leq x\leq 1$, we can obtain that 
\begin{equation}
\begin{split}
    &\vert K_t(x,y,t)\vert\leq 
    \frac{D^2}{2\varepsilon}\sum_{n=0}^{\infty}\frac{1}{n!}\Big(\frac{(x^2-y^2)D}{4\varepsilon}\Big)^n\\&\quad+\frac{D^2}{2\varepsilon}\sum_{n=0}^{\infty}\frac{n+2}{n!}\Big(\frac{(x^2-y^2)D}{4\varepsilon}\Big)^n\\&\leq  \frac{3D^2}{2\varepsilon}\sum_{n=0}^{\infty}\frac{1}{n!}\Big(\frac{(x^2-y^2)D}{4\varepsilon}\Big)^n\\&\quad+ \frac{D^2}{2\varepsilon}\frac{(x^2-y^2)D}{4\varepsilon}\sum_{n=1}^{\infty}\frac{1}{(n-1)!}\Big(\frac{(x^2-y^2)D}{4\varepsilon}\Big)^{n-1}\\&\leq \frac{3D^2}{2\varepsilon} e^{\frac{(x^2-y^2)D}{4\varepsilon}}+\frac{D^3(x^2-y^2)}{8\varepsilon^2}e^{\frac{(x^2-y^2)D}{4\varepsilon}}\\&\leq
    \frac{D^2}{2\varepsilon}\Big(3+\frac{D}{4\varepsilon}\Big) e^{\frac{D}{4\varepsilon}}.
\end{split}
\end{equation}
Differentiating \eqref{exK} with respect to $x$, we can obtain that
\begin{equation}
\begin{split}
    &K_x(x,y,t)\\&=-\frac{1}{2}e^{-\int_{0}^t\lambda(\xi)d\xi}\sum_{n=0}^{\infty}\frac{1}{n!(n+1)!}\Big(\frac{x^2-y^2}{4\varepsilon}\Big)^n F^{(n)}(t)\\&-\frac{x^2}{4\varepsilon}e^{-\int_{0}^t\lambda(\xi)d\xi}\sum_{n=1}^{\infty}\frac{1}{(n-1)!(n+1)!}\Big(\frac{x^2-y^2}{4\varepsilon}\Big)^{n-1}\\& \qquad\qquad\qquad\qquad\quad\times F^{(n)}(t).
\end{split}
\end{equation}
Using Assumption \ref{ass1}, relation \eqref{vvm} in Lemma \ref{ddf1}, and the fact that $0\leq y\leq x\leq 1$, we can obtain that 
\begin{equation}
    \begin{split}
        \vert K_x(x,y,t)\vert \leq &\frac{D}{2\varepsilon}\sum_{n=0}^{\infty}\frac{1}{n!}\Big(\frac{(x^2-y^2)D}{4\varepsilon}\Big)^n\\&+\frac{D^2}{4\varepsilon^2}\sum_{n=1}^{\infty}\frac{1}{(n-1)!}\Big(\frac{(x^2-y^2)D}{4\varepsilon}\Big)^{n-1}\\\leq & \frac{D}{2\varepsilon}e^{\frac{(x^2-y^2)D}{4\varepsilon}}+\frac{D^2}{4\varepsilon^2}e^{\frac{(x^2-y^2)D}{4\varepsilon}}
        \\\leq& \frac{D}{2\varepsilon}\Big(1+\frac{D}{2\varepsilon}\Big)e^{\frac{D}{4\varepsilon}}.
    \end{split}
\end{equation}\hfill$\qed$

\medskip

\begin{rmk}\label{nnba}
    For the solution to the gain kernel equations \eqref{ctcke11}-\eqref{eew11} given by \eqref{exK1}, we can obtain bounds similar to \eqref{ww1}-\eqref{ww8}. However, we only require to know the bound for $\vert L(x,y,t)\vert$ for $0\leq y\leq x\leq 1$ and $t\geq 0$. Similar to \eqref{ww1}, we can show that $L(x,y,t)$ satisfies
    \begin{equation}
        \vert L(x,y,t)\vert \leq \frac{D}{2\varepsilon} e^{\frac{D}{4\varepsilon}},
    \end{equation}
    for all $0\leq y\leq x\leq 1$ and $t\geq 0$. 
\end{rmk}

\medskip

\begin{lem}
For $d(t)$ given by \eqref{ffty}, it holds that\begin{equation}\label{ghm}
\dot{d}^{2}(t)\leq \rho_{1} d^{2}(t)+\alpha_{1}\Vert u[t]\Vert^{2}+\alpha_{2}u(1,t)\vert^{2},
\end{equation}
for some $\rho_{1},\alpha_{1},\alpha_{2}>0,$ for all $t\in(t_{j},t_{j+1}),j\in\mathbb{N}.$\end{lem}
\textit{Proof:} Taking the time derivative of \eqref{ffty} between $t\in(t_j,t_{j+1}),j\in\mathbb{N}$ along the solution of \eqref{ctpe1},\eqref{ctpe2},\eqref{mctpe3},\eqref{obetbc4} and using integration by parts twice, we can obtain that
\begin{equation}\label{nnkl}
\begin{split}
    &\dot{d}(t)= -\int_{0}^{1} k_t(y,t)u(y,t)dy-\int_{0}^{1}k(y,t)u_t(y,t)dy\\&=-\int_{0}^{1} k_t(y,t)u(y,t)dy-\lambda(t)\int_{0}^{1}k(y,t)u(y,t)dy\\&\quad-\varepsilon k(1,t)u_x(1,t)+\varepsilon k(0,t)u_x(0,t)+\varepsilon k_y(1,t)u(1,t)\\&\quad-\varepsilon k_y(0,t)u(0,t)-\varepsilon\int_{0}^{1}k_{yy}(y,t)u(y,t)dy.
\end{split}
\end{equation}
Note that we can show from \eqref{exK} and \eqref{kyt} that $k_y(0,t)=0$. Further note that $u_x(0,t)=0$ and 
\begin{equation}
\begin{split}
    u_x(1,t)&=-qu(1,t)+U_j\\&=-qu(1,t)+d(t)+\int_{0}^{1}k(y,t)u(y,t)dy
\end{split}
\end{equation}
from \eqref{mctpe3} and \eqref{obetbc4}. Thus, we can rewrite \eqref{nnkl} as
\begin{equation}
\begin{split}
    &\dot{d}(t)=-\varepsilon k(1,t)d(t)+\varepsilon \big(qk(1,t)+k_y(1,t)\big)u(1,t)\\&-\int_{0}^{1}\Big(k_t(y,t)+\lambda(t)k(y,t)+\varepsilon k(1,t)k(y,t)\\&\qquad\qquad\qquad\qquad\qquad+\varepsilon k_{yy}(y,t)\big)\Big)u(y,t)dy,
\end{split}
\end{equation}
from which we can obtain using Assumption \ref{ass1} and Cauchy-Schwarz inequality that
\begin{equation}\label{dnot}
    \begin{split}
        &\vert\dot{d}(t)\vert\\&\leq \varepsilon\max_{t\geq 0}\big(\vert k(1,t)\vert\big)\vert d(t)\vert       +\varepsilon\Big(q\max_{t\geq 0}\vert k(1,t)\vert\\&\qquad\qquad+\max_{t\geq 0}\vert k_y(1,t)\vert\Big)\vert u(1,t)\vert\\&+\Big(\max_{t\geq 0,y\in[0,1]}\vert k_{t}(y,t)\vert+D\max_{t\geq 0,y\in[0,1]}\vert k(y,t)\vert\\&\qquad+\varepsilon\max_{t\geq 0}\vert k(1,t)\vert\max_{t\geq 0, y\in[0,1]}\vert k(y,t)\vert\\&\qquad+\varepsilon\max_{t\geq 0,y\in[0,1]}\vert k_{yy}(y,t)\vert\Big)\Vert u[t]\Vert.
    \end{split}
\end{equation}
As $r(t)=q-\frac{\lambda(t)}{2\varepsilon}$, using Assumption \ref{ass1}, we can show that
\begin{equation}\label{rr1}
    \vert r(t)\vert\leq q+\frac{D}{2\varepsilon},
\end{equation}
and 
\begin{equation}\label{rr2}
    \vert\dot{r}(t)\vert\leq \frac{D^2}{2\varepsilon}.
\end{equation}
Considering \eqref{kyt},\eqref{ww1}-\eqref{ww8},\eqref{rr1},\eqref{rr2}, we can obtain the following relations:

\begin{equation}\label{yy1}
\begin{split}
    &\vert k_t(y,t)\vert \leq \vert K_{xt}(1,y,t)\vert+\vert\dot{r}(t)\vert\cdot\vert K(1,y,t)\vert\\&\qquad\qquad\qquad\qquad\qquad\qquad+\vert r(t)\vert\cdot\vert K_t(1,y,t)\vert\\&\leq \vert K_{xt}(1,y,t)\vert+\frac{D^2}{2\varepsilon}\vert K(1,y,t)\vert+\Big(q+\frac{D}{2\varepsilon}\Big)\vert K_t(1,y,t)\vert \\&\leq \max_{0 \leq y\leq  x\leq 1}\Big(\vert K_{xt}(x,y,t)\vert+\frac{D^2}{2\varepsilon}\vert K(x,y,t)\vert\\&\qquad\qquad\qquad\qquad\qquad\qquad+\Big(q+\frac{D}{2\varepsilon}\Big)\vert K_t(x,y,t)\vert\Big)\\&\leq \frac{D^2}{\varepsilon}\Big(\frac{3}{2}+\frac{3q}{2}+\frac{qD}{8\varepsilon}+\frac{17D}{8\varepsilon}+\frac{D^2}{8\varepsilon^2}\Big)e^{\frac{D}{4\varepsilon}},
\end{split}
\end{equation}

\begin{equation}
\begin{split}
    &\vert k(y,t)\vert\leq \vert K_x(1,y,t)\vert +\vert r(t)\vert\cdot\vert K(1,y,t)\vert\\&\leq \max_{0 \leq y\leq  x\leq 1}\Big(\vert K_{x}(x,y,t)\vert+\Big( q+\frac{D}{2\varepsilon}\Big)\vert K(x,y,t)\vert\Big)\\&\leq 
    \frac{D}{2\varepsilon}\Big(1+q+\frac{D}{\varepsilon}\Big)e^{\frac{D}{4\varepsilon}},
\end{split}
\end{equation}

\begin{equation}
\begin{split}
    &\vert k_y(y,t)\vert\leq \vert K_{xy}(1,y,t)\vert +\vert r(t)\vert\cdot \vert K_y(1,y,t)\vert\\&\leq \max_{0\leq y\leq x\leq 1}\Big(\vert K_{xy}(x,y,t)\vert +\Big( q+\frac{D}{2\varepsilon}\Big)\vert K_y(x,y,t)\vert\Big)\\&\leq \frac{D^2}{4\varepsilon^2} \Big(1+q+\frac{D}{\varepsilon}\Big) e^{\frac{D}{4\varepsilon}},
\end{split}
\end{equation}
and
\begin{equation}\label{yy4}
\begin{split}
    &\vert k_{yy}(y,t)\vert\leq \vert K_{xyy}(1,y,t)\vert+\vert r(t)\vert\cdot\vert K_{yy}(1,y,t)\vert\\&\leq \max_{0\leq y\leq x\leq 1}\Big(\vert K_{xyy}(x,y,t)\vert+\Big( q+\frac{D}{2\varepsilon}\Big)\vert K_{yy}(x,y,t)\vert\Big)\\&\leq 
    \frac{D^2}{4\varepsilon^2}\Big(1+q+\frac{qD}{2\varepsilon}+\frac{D}{\varepsilon}+\frac{D^2}{2\varepsilon^2}\Big)e^{\frac{D}{4\varepsilon}}.
\end{split}
\end{equation}
Using Young's inequality and \eqref{yy1}-\eqref{yy4}, we can obtain from \eqref{dnot} that
\begin{equation}
    \dot{d}^2(t)\leq \rho_1 d^2(t)+\alpha_1\Vert u[t]\Vert^2+\alpha_2 u^2(1,t)
\end{equation}  
where
\begin{equation}
    \rho_1 = \frac{3D^2}{4}\Big(1+q+\frac{D}{\varepsilon}\Big)^2 e^{\frac{D}{2\varepsilon}},
\end{equation}

\begin{equation}\label{al1}
    \begin{split}&\alpha_1= \frac{3D^4}{\varepsilon^2}\bigg(\frac{9}{4}+\frac{9q}{4}+\frac{4D}{\varepsilon}+\frac{9D}{8\varepsilon}+\frac{D^2}{4\varepsilon^2}\\&\qquad\qquad\qquad\qquad+\frac{1}{4}\Big(1+q+\frac{D}{\varepsilon}\Big)^2 e^{\frac{D}{4\varepsilon}}\bigg)^2\ e^{\frac{D}{2\varepsilon}},
    \end{split}
\end{equation}

\begin{equation}\label{al2}
    \alpha_2=\frac{3D^2}{4}\Big(q+\frac{D}{2\varepsilon}\Big)^2\Big(1+q+\frac{D}{\varepsilon}\Big)^2e^{\frac{D}{2\varepsilon}}.
\end{equation}
This completes the proof. \hfill$\qed$

\medskip

\begin{thme}\label{kkh}Under the event-triggered boundary control in Definition \ref{def1}, with $\beta_{1},\beta_{2}$ chosen as\begin{equation}\label{betas}
\beta_{1}=\frac{\alpha_{1}}{\gamma(1-\sigma)},\hspace{5pt}\beta_{2}=\frac{\alpha_{2}}{\gamma(1-\sigma)},
\end{equation}where $\alpha_{1},\alpha_{2}$ given by \eqref{al1},\eqref{al2} and $\sigma\in(0,1)$,  there exists a minimal dwell-time $\tau>0$ between two triggering times, \textit{i.e.,} there exists a constant $\tau>0$ such that $t_{j+1}-t_{j}\geq\tau,$ for all $j\in\mathbb{N}$, which is independent of the initial conditions and only depends on the system and control parameters.\end{thme}

\textit{Proof:}  The proof is very similar to that of Theorem 1 in \cite{rathnayake2021observer}, and hence omitted. 

\medskip

\begin{cllry}\label{corf}
Under Assumption \ref{ass1}, for every $u[0]\in L^{2}(0,1)$, there exist unique solution $u:\mathbb{R}_+\times[0,1]\rightarrow\mathbb{R}$ such that $u\in C^0(\mathbb{R}_+;L^2(0,1)\cap C^1(J\times [0,1])$ with $u[t]\in C^2([0,1])$ which satisfy \eqref{ctpe2},\eqref{mctpe3} all $t>0$ and \eqref{ctpe1} for all $t>0,x\in(0,1),$ where $J=\mathbb{R_{+}}\text{\textbackslash}\{t_{j}\geq 0,j\in\mathbb{N}\}$. The increasing sequence $\{t_{j}\geq 0,j\in\mathbb{N}\}$ is determined by the set of rules  given in Definition \ref{def1}.
\end{cllry}

\textit{Proof:} This is a straightforward consequence of Theorem \ref{thm11} and \ref{kkh}. The solutions are constructed
iteratively between successive triggering times. \hfill $\qed$ 

\medskip

\begin{thme}\label{thm2}Let $\gamma,\eta>0$  be design parameters, and $\beta_{1},\beta_{2}$ are chosen according to \eqref{betas}.  Subject to Assumption \ref{ass2}, the parameter $\rho>0$ is set as 
\begin{equation}\label{hhjk}
\rho=\frac{\varepsilon\kappa B}{2},
\end{equation}
for $B,\kappa>0$ are chosen such that
\begin{equation}\label{Bs}
\begin{split}
B\bigg(\varepsilon\min\Big\{q-\frac{D}{2\varepsilon}-\frac{1}{2},\frac{1}{2}\Big\}&-\frac{\varepsilon}{2\kappa}\bigg)-2\beta_1\Big(1+\frac{D}{2\varepsilon}e^{\frac{D}{4\varepsilon}}\Big)^2\\&-2\beta_2-\frac{\beta_2D^2}{\varepsilon^2}e^{\frac{D}{2\varepsilon}}>0.
\end{split}
\end{equation} Note from Assumption \ref{ass2} that $q-D/2\varepsilon>1/2$. Then, under Assumption \ref{ass1} and \ref{ass2}, the closed-loop system which consists of the plant and the  event-triggered boundary controller \eqref{obetbc1}-\eqref{obetbc4} has a unique solution and globally exponentially converges to zero, \text{i.e.,} $\Vert u[t]\Vert\rightarrow 0$ as $t\rightarrow \infty.$\end{thme}

\textit{Proof:} From Corollary \ref{corf}, the existence and the uniqueness of solutions to the system \eqref{ctpe1},\eqref{ctpe2},\eqref{mctpe3} under the event-triggered boundary control in Definition \ref{def1} is guaranteed. Now let us show that the closed-
loop system is globally $L^2$-exponentially convergent to zero.

Let us choose the following candidate Lyapunov function
noting that $m(t)>0$ for all $t\geq 0$:
\begin{equation}\label{jjl}
    V = \frac{B}{2}\int_{0}^{1}w^2(x,t)dx+m(t),
\end{equation}
where $B>0$ is chosen to satisfy \eqref{Bs}. Differentiating \eqref{jjl} along the solution of \eqref{ettsm1}-\eqref{ettsm3} and \eqref{obetbc3}, and using integration by parts twice, we can obtain that
\begin{equation}\label{ddf}
    \begin{split}
        &\dot{V}(t)=-\varepsilon Br(t) w^2(1,t)+\varepsilon B d(t) w(1,t)\\&-\varepsilon B\Vert w_x[t]\Vert^2-\eta m(t)-\rho d^2(t)+\beta_1\Vert u[t]\Vert^2+\beta_2 u^2(1,t).
    \end{split}
\end{equation}
From Young's inequality, we can write that
\begin{equation}\label{ss1}
    \varepsilon B d(t)w(1,t)=\frac{\varepsilon B}{2\kappa} w^2(1,t)+\frac{\varepsilon\kappa B}{2}d^2(t),
\end{equation}
for $\kappa>0$ chosen to satisfy \eqref{Bs}. Furthermore, from Poincare's inequality, we have that
\begin{equation}\label{ss2}
    \Vert w[t]\Vert^2\leq 2 w^2(1,t)+4\Vert w_x[t]\Vert^2.
\end{equation}
Thus, using \eqref{ss1},\eqref{ss2}, we can obtain from \eqref{ddf} that
\begin{equation}\label{nnm}
\begin{split}
    &\dot{V}\leq -\varepsilon Br(t)w^2(1,t)+\frac{\varepsilon B}{2\kappa} w^2(1,t)+\frac{\varepsilon B \kappa}{2}d^2(t)\\&
    +\varepsilon B\Big(\frac{1}{2}w^2(1,t)-\frac{1}{4}\Vert w[t]\Vert^2\Big)-\eta m(t)-\rho d^2(t)\\&+\beta_1\Vert u[t]\Vert^2+\beta_2 u^2(1,t).
\end{split}
\end{equation}
Recalling \eqref{hhjk}, we can rewrite \eqref{nnm} as 
\begin{equation}\label{hhn}
\begin{split}
    &\dot{V}\leq -\varepsilon B \Big(\big(r(t)-\frac{1}{2}\big)-\frac{1}{2\kappa}\Big)w^2(1,t)-\frac{\varepsilon B}{4}\Vert w[t]\Vert^2\\&-\eta m(t)+\beta_1\Vert u[t]\Vert^2+\beta_2 u^2(1,t).
\end{split}
\end{equation}
Using Young's inequality and Cauchy-Schwarz inequality, we can obtain from \eqref{ink} that
\begin{equation}
\begin{split}
    u^2(1,t)&\leq 2w^2(1,t)+2\int_{0}^{1}L^2(1,y,t)dy\Vert w[t]\Vert^2,
\end{split}
\end{equation}
and 
\begin{equation}\label{yc1}
\Vert u[t]\Vert^2\leq \bigg(1+\Big(\int_{0}^{1}\int_{0}^{x}L^{2}(x,y)dydx\Big)^{1/2}\bigg)^2\Vert w[t]\Vert^2. 
\end{equation}
Thus, recalling Remark \ref{nnba}, we can show that 
\begin{equation}
    u^2(1,t)=2w^2(1,t)+\frac{D^2}{2\varepsilon^2}e^{\frac{D}{2\varepsilon}}\Vert w[t]\Vert^2,
\end{equation}
and
\begin{equation}\label{hhbn}
    \begin{split}
\Vert u[t]\Vert^2\leq \bigg(1+\frac{D}{2\varepsilon}e^{\frac{D}{4\varepsilon}}\bigg)^2\Vert w[t]\Vert^2.
    \end{split}
\end{equation}
Therefore, we can obtain from \eqref{hhn} as
\begin{equation}\label{ppn}
\begin{split}
    &\dot{V}\leq -\bigg(\varepsilon B\Big(\big(r(t)-\frac{1}{2}\big)-\frac{1}{2\kappa}\Big)-2\beta_2\bigg)w^2(1,t)\\&-\bigg(\frac{\varepsilon B}{4}-\beta_1\Big(1+\frac{D}{2\varepsilon}e^{\frac{D}{4\varepsilon}}\Big)-\frac{\beta_2 D^2}{2\varepsilon^2}e^{\frac{D}{2\varepsilon}}\bigg)\Vert w[t]\Vert^2-\eta m(t).
\end{split}
\end{equation}
Considering Assumption \ref{ass1} and \ref{ass2}, and equation \eqref{RR}, we can write that 
\begin{equation}
    r(t)\geq q-\frac{D}{2\varepsilon}>\frac{1}{2}.
\end{equation}
Thus, we can rewrite \eqref{ppn} as
\begin{equation}\label{nnb}
\begin{split}
    &\dot{V}\leq -\bigg(\varepsilon B\Big(\big(q-\frac{D}{2\varepsilon}-\frac{1}{2}\big)-\frac{1}{2\kappa}\Big)-2\beta_2\bigg)w^2(1,t)\\&-\bigg(\frac{\varepsilon B}{4}-\beta_1\Big(1+\frac{D}{2\varepsilon}e^{\frac{D}{4\varepsilon}}\Big)-\frac{\beta_2 D^2}{2\varepsilon^2}e^{\frac{D}{2\varepsilon}}\bigg)^2\Vert w[t]\Vert^2-\eta m(t).
\end{split}
\end{equation}
Let
\begin{equation}
    b_1:=\varepsilon B\Big(\big(q-\frac{D}{2\varepsilon}-\frac{1}{2}\big)-\frac{1}{2\kappa}\Big)-2\beta_2,
\end{equation}
and 
\begin{equation}
    b_2:=\frac{\varepsilon B}{4}-\beta_1\Big(1+\frac{D}{2\varepsilon}e^{\frac{D}{4\varepsilon}}\Big)^2-\frac{\beta_2 D^2}{2\varepsilon^2}e^{\frac{D}{2\varepsilon}}.
\end{equation}
As $B>0$ and $\kappa>0$ have been selected to satisfy \eqref{Bs}, we can observe that $b_1,b_2>0$. Thus, considering \eqref{jjl}, we can obtain from \eqref{nnb} that   
\begin{equation}
    \dot{V}\leq -2\varrho V,
\end{equation}
where
\begin{equation}
    \varrho = \min\Big\{\frac{b_2}{B},\frac{\eta}{2}\Big\},
\end{equation}
from which we can obtain using standard arguments that  \begin{equation}
    V(t)\leq e^{-2\varrho t}V(0).
\end{equation}
This implies due to \eqref{jjl} that $\Vert w[t]\Vert\rightarrow 0$ as $t\rightarrow \infty$. Therefore, considering \eqref{hhbn}, we can conclude that $\Vert u[t]\Vert\rightarrow 0$ as $t\rightarrow \infty$.  This completes the proof. \hfill$\qed$ 
\section{Numerical Simulations}
We consider a reaction-diffusion system with $\varepsilon=1,\lambda(t)=3/(1+t), t\geq 0,q=3,$ and the initial conditions $u[0]=10x^2(x-1)^{2}$. It is easy to show that $\lambda(t)$ satisfies Assumption \ref{ass1} with $D=3$. The parameters for the event-trigger in Definition \ref{def1} are chosen as follows: $m(0)=10^{-4},\gamma=1,\eta=1,$ and  $\sigma=0.5$. It can be shown using \eqref{al1},\eqref{al2} that $\alpha_{1}=3.0084\times 10^6, \alpha_2= 3.9624\times 10^3$. Therefore, from \eqref{betas}, we can obtain $\beta_{1}=6.0167\times 10^6;\beta_{2}=7.9248\times 10^3$. Let us choose $B$ and $\kappa$ as $B=8.4054\times 10^8$ and $\kappa=2$ so that \eqref{Bs} is satisfied. Then, from \eqref{hhjk}, we can obtain $\rho=8.4054\times 10^8$.

Fig. 2 illustrates the evolution of event-triggered control (ETC) and continuous-time  control (CTC) inputs. We can clearly observe that ETC significantly reduces the number of control updates compared to CTC. Fig. 3 shows the evolution of $\Vert u[t]\Vert$, and we can observe that the convergence under ETC is the fastest.   

\begin{figure}
\centering
\includegraphics[scale=0.095]{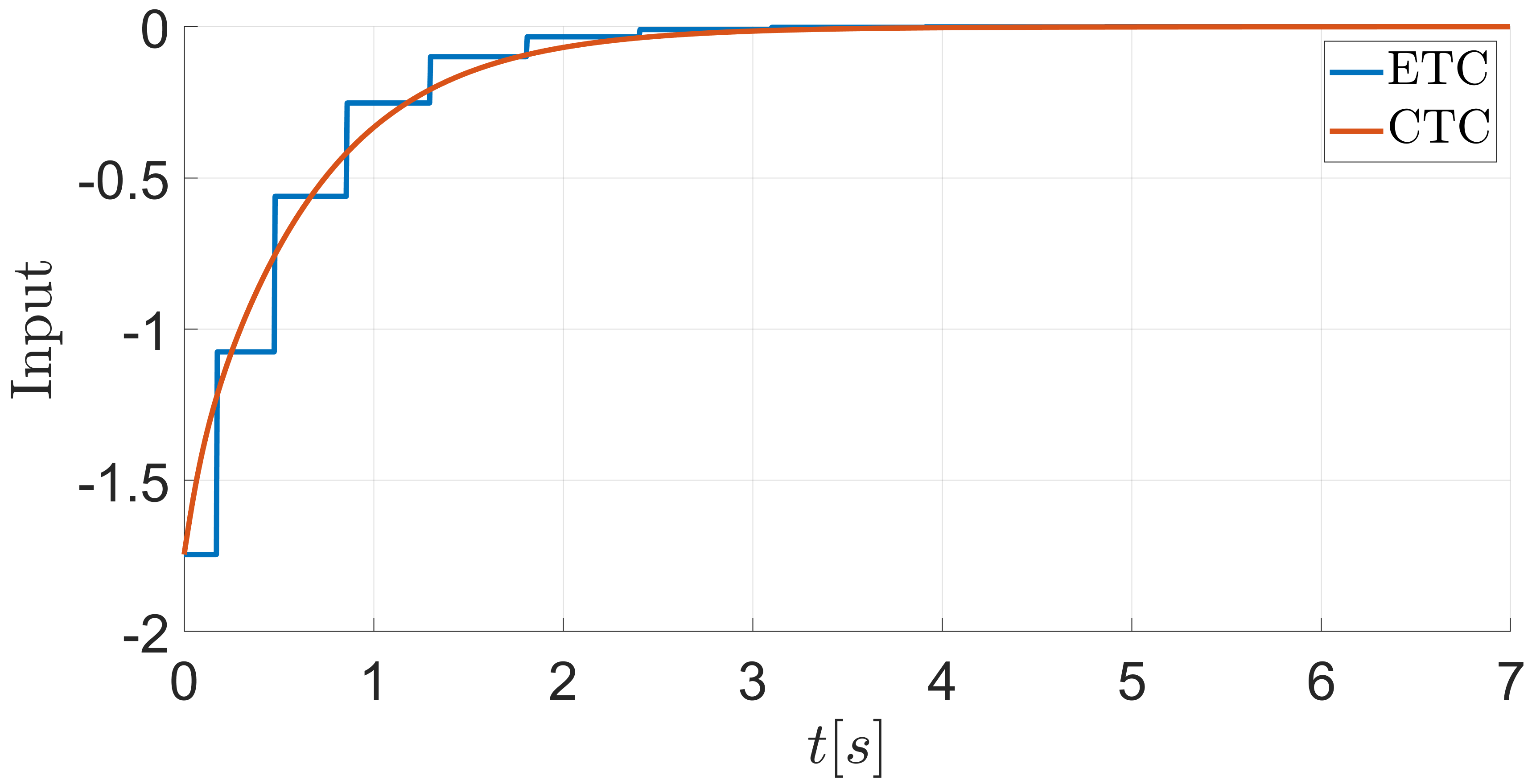}
\caption{Event-triggered control (ETC) and continuous-time control (CTC) inputs.}
\end{figure}

\begin{figure}
\centering
\includegraphics[scale=0.095]{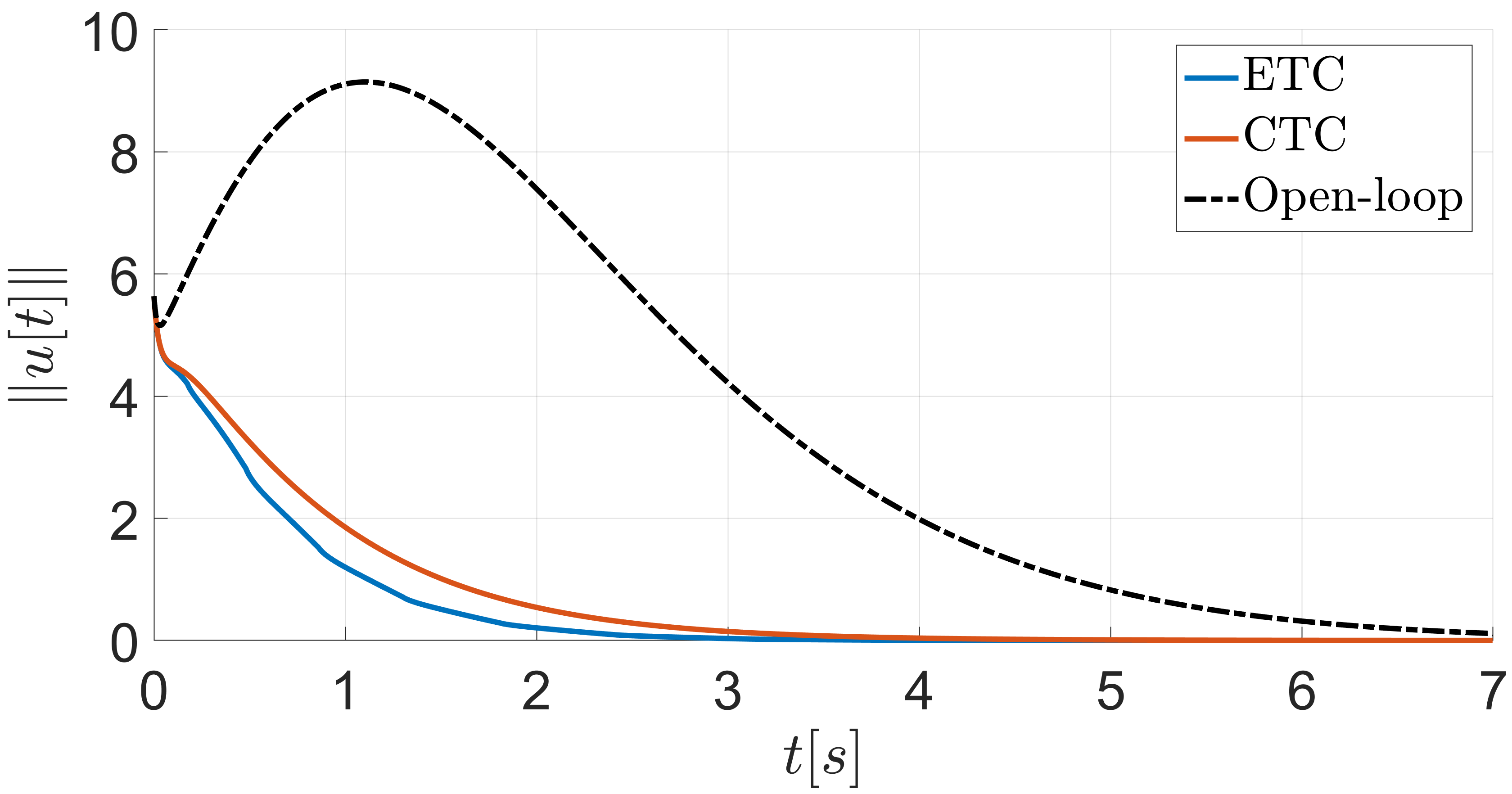}
\caption{Evolution of $\Vert u[t]\Vert$.}
\end{figure}

\section{Conclusion}
This paper has proposed an event-triggered full-state feedback boundary control strategy for a class of reaction-diffusion systems with time-dependent reactivity under Robin boundary actuation. We have used a dynamic event triggering condition to determine when the control value needs to be updated. Under the proposed strategy, we have proved the existence of a uniform minimal-dwell time between two updates, which excludes Zeno behavior. Further, we have shown the well-posedness of the closed-loop system  and its global $L^{2}$-exponential convergence to zero.

\section*{Appendix A\\Solution to \eqref{ctck} and \eqref{ctck2}}
Let us make the following change of variables
\begin{equation}\label{ggn1}
    K(x,y,t)=-\frac{x}{2}e^{-\int_{0}^{t}\lambda(\xi)d\xi}f(z,t), \text{ }z=\sqrt{x^2-y^2},
\end{equation}
from which we get the following PDE in one spatial variable for the
function $f(z,t)$

\begin{equation}
    f_t(x,t)=\varepsilon f_{zz}(z,t)+\frac{3\varepsilon}{z}f_{z}(z,t),
\end{equation}
with the boundary conditions 
\begin{align}
    f_z(0,t)&=0,\\
    f(0,t)&=\frac{\lambda(t)}{\varepsilon} e^{\int_{0}^{t}\lambda(\xi)d\xi}.
\end{align}
We further make the following change of coordinates and variables
\begin{equation}
    \bar{t}=\varepsilon t,
\end{equation}
and 
\begin{equation}
    \bar{f}(z,\bar{t})= f(z,t)
\end{equation}
from which we obtain the following PDE
\begin{align}\label{ee1}
    \bar{f}_{\bar{t}}(z,\bar{t})&=\bar{f}_{zz}(z,\bar{t})+\frac{3}{z}\bar{f}_{z}(z,\bar{t})\\\label{ee2}
    \bar{f}_z(0,\bar{t})&=0\\\label{ee3}
    \bar{f}(0,\bar{t})&=\frac{\lambda(\bar{t}/\varepsilon)}{\varepsilon} e^{\int_{0}^{\bar{t}/\varepsilon}\lambda(\xi)d\xi}:=F(\bar{t}/\varepsilon)=F(t).
\end{align}
The unique solution to \eqref{ee1}-\eqref{ee3} is given by \cite{polyanin2001handbook}
\begin{equation}\label{ggn2}
\begin{split}
\bar{f}(z,\bar{t})=f(z,t)=&\sum_{n=0}^{\infty}\frac{1}{n!(n+1)!}\Big(\frac{z^2}{4}\Big)^n\frac{d^n F(t)}{d\bar{t}^n}\\=&
\sum_{n=0}^{\infty}\frac{1}{n!(n+1)!}\Big(\frac{z^2}{4\varepsilon}\Big)^n F^{(n)}(t). 
\end{split}
\end{equation}
Thus, considering \eqref{ggn1} and \eqref{ggn2}, we can obtain the closed-form solution \eqref{exK} for all $0\leq y\leq x\leq 1$ and $t\geq 0$.

By comparing the PDEs \eqref{ctcke1}-\eqref{eew} and \eqref{ctcke11}-\eqref{eew11}, we can observe that
\begin{equation}
    L\big(x,y,t;\lambda(t)\big)=-K\big(x,y,t;-\lambda(t)\big),
\end{equation}
from which we can show that the solution for $L(x,y,t)$ for all $0\leq y\leq x\leq 1$ and $t\geq 0$ is given by \eqref{exK1}.

\bibliographystyle{IEEEtran}
\bibliography{main}
\end{document}